\documentclass[twocolumn,preprintnumbers,amsmath,amssymb]{revtex4}

\usepackage{epsfig}
\usepackage{subfigure}
\usepackage{graphics}
\usepackage{amsmath}

\begin{document}

\title{Shear and bulk viscosities of strongly interacting ``infinite'' parton-hadron matter within
the parton-hadron-string dynamics transport approach}

\author{V.~Ozvenchuk}

\email{ozvenchuk@fias.uni-frankfurt.de}

\affiliation{%
 Frankfurt Institute for Advanced Studies, %
 60438 Frankfurt am Main, %
 Germany %
}

\author{O.~Linnyk}%
\affiliation{%
 Institut f\"ur Theoretische Physik, %
  Universit\"at Giessen, %
  35392 Giessen, %
  Germany %
}

\author{M.~I.~Gorenstein}%
\affiliation{%
 Bogolyubov Institute for Theoretical Physics, %
 Kiev, %
 Ukraine %
}

\author{E.~L.~Bratkovskaya}%
\affiliation{%
 Institut f\"ur Theoretische Physik, %
 Johann Wolfgang Goethe-Universit\"at, %
 60438 Frankfurt am Main, %
 Germany; %
 Frankfurt Institute for Advanced Studies %
 60438 Frankfurt am Main, %
 Germany %
}
\author{W.~Cassing}
\affiliation{%
  Institut f{\"u}r Theoretische Physik, %
  Universit\"at Giessen, %
  35392 Giessen, %
  Germany %
}

\date{\today}

\begin{abstract}
We study the shear and bulk viscosities of partonic and hadronic
matter as functions of temperature $T$ within the
parton-hadron-string dynamics (PHSD) off-shell transport approach.
Dynamical hadronic and partonic systems in equilibrium are studied
by the PHSD simulations in a finite box with periodic boundary
conditions. The ratio of the shear viscosity to entropy density
$\eta (T)/s (T)$ from PHSD shows a minimum (with a value of about $
0.1$) close to the critical temperature $T_c$, while it approaches
the perturbative QCD limit at higher temperatures in line with
lattice QCD (lQCD) results. For $T < T_c$, i.e., in the hadronic
phase, the ratio $\eta/s$ rises fast with decreasing temperature due
to a strong decrease of the entropy density $s$ in the hadronic
phase at decreasing $T$. Within statistics, we obtain practically
the same results in the Kubo formalism and in the relaxation time
approximation. The bulk viscosity $\zeta (T)$---evaluated in the
relaxation time approach---is found to strongly depend on the
effects of mean fields (or potentials) in the partonic phase. We
find a significant rise of the ratio $\zeta (T)/s (T)$ in the
vicinity of the critical temperature $T_c$, when consistently
including the scalar mean-field from PHSD, which is also in
agreement with that from lQCD calculations. Furthermore, we present
the results for the ratio $(\eta + 3 \zeta/4)/s$, which is found to
depend nontrivially on temperature and to generally agree with the
lQCD calculations as well. Within the PHSD calculations, the strong
maximum of $\zeta(T)/\eta (T)$ close to $T_c$ has to be attributed
to mean-field (or potential) effects that in PHSD are encoded in the
temperature dependence of the quasiparticle masses, which is related
to the infrared enhancement of the resummed (effective) coupling
$g(T)$.
\end{abstract}

\maketitle

\section{Introduction}
High-energy heavy-ion reactions are studied experimentally and
theoretically to obtain information about the properties of nuclear
matter under the extreme conditions of high baryon density and/or
temperature. Ultrarelativistic heavy-ion collisions at the
Relativistic Heavy-Ion Collider (RHIC) and the Large Hadron Collider
(LHC) at CERN have produced a new state of matter, the quark-gluon
plasma (QGP), for a couple of fm/$c$. The produced QGP shows
features of a strongly interacting fluid unlike a weakly interacting
parton gas \cite{StrCoupled1,StrCoupled2,StrCoupled3,Peshier}. Large
values of the observed azimuthal asymmetry of charged particles in
momentum space, i.e., the elliptic flow $v_2$
\cite{STAR,PHENIX,BRAHMS,PHOBOS,ALICE}, could quantitatively be well
described by  hydrodynamics up to transverse momenta on the order of
$1.5$~GeV/$c$
\cite{IdealHydro1,IdealHydro2,IdealHydro3,IdealHydro4,IdealHydro5,IdealHydro6}.
A perfect fluid has been defined as  having a zero shear viscosity,
$\eta$; yet semiclassical arguments have been given suggesting that
the shear viscosity cannot be arbitrarily small
\cite{NonzeroViscosity}. Indeed, the lower bound for the shear
viscosity to entropy density ratio $\eta/s=1/4\pi$ was conjectured
by Kovtun-Son-Starinets (KSS) \cite{KSS} for infinitely coupled
supersymmetric Yang-Mills gauge theory based on the anti de
Sitter/conformal field theory (AdS/CFT) duality conjecture. On the
basis of holographically dual computations \cite{Buchel}, also for
the bulk viscosity of strongly coupled gauge theory plasmas a lower
bound was conjectured: $\zeta/\eta\geqslant2(1/3-c_s^2)$, where
$c_s$ is the speed of sound. Empirically, relativistic viscous
hydrodynamic calculations---using the Israel-Stewart
framework---require a very small but finite $\eta/s$ of $0.08-0.24$
in order to reproduce the RHIC elliptic flow $v_2$ data
\cite{ViscousHydro1,ViscousHydro2,ViscousHydro3,ViscousHydro4}. The
main uncertainty in these estimates results from the equation of
state and the initial conditions employed in the hydrodynamical
calculations as well as in the temperature dependence of
$\eta/s(T)$.

Thus not only the absolute values of the shear and bulk viscosities
are of great interest but also the temperature dependence of these
coefficients, which is expected to be quite strong. There is
evidence from atomic and molecular systems that $\eta/s$ should have
a minimum in the vicinity of the phase transition or---in case of
strong interactions at vanishing chemical potential---of the rapid
crossover between hadronic matter and the quark-gluon plasma
\cite{Minshear,Mattiello,Toneev}. Furthermore, it is argued that the
ratio of the bulk viscosity to entropy density $\zeta/s$ should have
a maximum close to $T_c$---as suggested by lattice QCD---and might
even diverge in the case of a second-order phase transition
\cite{MaxBulk1,MaxBulk2,MaxBulk3,MaxBulk4,MaxBulk5,Pratt}. Such a
peak in the bulk viscosity can lead to instabilities in viscous
hydrodynamics simulations for heavy-ion collisions and possibly to
clusterization effects~\cite{Giorgio}.

Shear and bulk viscosities of strongly interacting systems have been
evaluated within different approaches. Calculations have been
performed at  high temperatures, where perturbation theory can be
applied \cite{pQCD1,pQCD2}, as well as at extremely low temperatures
\cite{pQCD2,pQCD3,pQCD4}. First results for shear and bulk
viscosities obtained within lattice QCD (lQCD) simulations above the
critical temperature of pure gluon matter have been presented in
Refs.~\cite{lQCDtransport1,lQCDtransport2,lQCDtransport3,lQCDtransport4}.
There are several methods for the calculation of shear and bulk
viscosities for strongly interacting systems: the relaxation time
approximation (RTA) \cite{RTA}, the Chapmann-Enskog (CE) method
\cite{CE}, and the Green-Kubo approach \cite{Green,Kubo}. The RTA
method has been  used to calculate the viscosity
\cite{NonzeroViscosity,MaxBulk5,Bluhm,GrecoQGP,Plumari,Thoma,Khvorostukhin1,Khvorostukhin2},
as well as the Green-Kubo approach
\cite{Peshier,Bass,BassQGP,Muronga,Wesp,Pal,GrecoQGP}, for both
hadronic and partonic matter providing a rough picture of the
transport properties of strongly interacting matter.

In this study we calculate the shear and bulk viscosities as a
function of temperature  (or energy density) with the
parton-hadron-string dynamics (PHSD) transport approach that has
provided a good description of collective flow properties and
differential particle spectra in nucleus-nucleus collisions from
lower CERN Super Proton Synchrotron (SPS) to RHIC energies
\cite{PHSD1,PHSD2,Olena,Olena2,Koncha11}. In this approach the shear
and bulk viscosities do not enter as external parameters but are
generic properties of the matter under consideration and can be
calculated for systems in equilibrium as a function of temperature
explicitly without incorporating any additional parameters.
Furthermore,  the PHSD aproach allows one to evaluate the transport
coefficients within the partonic phase as well as within the
hadronic phase on the same footing.

The paper is organized as follows. In Sec.~II we provide a brief
reminder of the off-shell dynamics and the ingredients of the PHSD
transport approach. We then first present in Sec.~III the actual
results for the shear and bulk  viscosities in  ``infinite''
parton-hadron matter within the PHSD employing the Green-Kubo
formalism and the RTA and compare these results to the available
lQCD results. The summary and conclusions are given in Sec.~IV.

\section{The PHSD transport approach}
In this work we extract the shear and bulk viscosities for
``infinite'' parton-hadron matter employing different methods within
the PHSD transport approach \cite{PHSD1,PHSD2}, which is based on
generalized transport equations on the basis of the off-shell
Kadanoff-Baym equations \cite{Kadanoff1,Kadanoff2} for Green's
functions in phase-space representation (in the first-order gradient
expansion, beyond the quasiparticle approximation). The approach
consistently describes the full evolution of a relativistic
heavy-ion collision from the initial hard scatterings and string
formation through the dynamical deconfinement phase transition to
the strongly interacting quark-gluon plasma (sQGP) as well as
hadronization and the subsequent interactions in the expanding
hadronic phase. In the hadronic sector PHSD is equivalent to the
hadron-string dynamics (HSD) transport approach
\cite{CBRep98,Brat97}---a covariant extension of the
Boltzmann-Uehling-Uhlenbeck approach \cite{Cass90}---that has been
used for the description of $pA$ and $AA$ collisions from GSI Heavy
Ion Synchrotron (SIS) to RHIC energies in the past. In PHSD the
partonic dynamics is based on the dynamical quasiparticle model
(DQPM) \cite{DQPM1,DQPM2,DQPM3}, which describes QCD properties in
terms of single-particle Green's functions (in the sense of a
two-particle irreducible approach) and reproduces lattice QCD
results---including the partonic equation of state---in
thermodynamic equilibrium.

\subsection{Reminder of the DQPM}
In the scope of the DQPM the running coupling constant $g^2$
(squared) for partons  is approximated (for $T>T_c$) by
\begin{equation}
g^2(T/T_{c})=\frac{48\pi^2}{(11N_{c}-2N_{f})\ln[\lambda^2(T/T_{c}-T_{s}/T_{c})^2]},
\label{running}
\end{equation}
where the parameters $\lambda=2.42$ and $T_{s}/T_{c}=0.56$ have been
extracted from a fit to the lattice data for purely gluonic systems
($N_f$=0) as described in Ref. \cite{DQPM3}. In Eq.~(\ref{running}),
$N_{c}=3$ stands for the number of colors, $T_c$ is the critical
temperature ($=158$ MeV), and $N_{f}$ denotes the number of flavors.
In the actual PHSD calculations for $N_f=3$ we employ a slightly
different analytical form for $g^2(T/T_c)$ that has been fitted to
the lattice data from Ref. \cite{lQCD}. For the details we refer the
reader to Ref. \cite{Bratkovskaya:2011wp}.

The functional forms for the dynamical quasiparticle masses (for
gluons and quarks) are chosen so that they become identical to the
perturbative thermal masses in the asymptotic high-temperature
regime; i.e., for gluons
\begin{equation}
M^2_{g}(T)=\frac{g^2(T/T_c)}{6}\left(\left(N_{c}+\frac{1}{2}N_{f}\right)T^2
+\frac{N_c}{2}\sum_{q}\frac{\mu^{2}_{q}}{\pi^2}\right)\ ,
\end{equation}
and for quarks (antiquarks)
\begin{equation} \label{Mq9}
M^2_{q(\bar
q)}(T)=\frac{N^{2}_{c}-1}{8N_{c}}g^2\left(T^2+\frac{\mu^{2}_{q}}{\pi^2}\right)\
,
\end{equation}
but the running coupling $g(T/T_c)$ is the resummed coupling of
Eq.~(1). The effective quarks, antiquarks, and gluons in the DQPM
have finite widths, which for $\mu_{q}=0$ are adopted in the
following form \cite{Pisarski}:
\begin{eqnarray}\label{widthg}
\Gamma_{g}(T)&=&\frac{1}{3}N_{c}\frac{g^2T}{8\pi}\ln\left(\frac{2c}{g^2}+1\right)\
,\\
\label{widthq}\Gamma_{q(\bar
q)}(T)&=&\frac{1}{3}\frac{N^{2}_{c}-1}{2N_{c}}\frac{g^2T}{8\pi}\ln\left(\frac{2c}{g^2}+1\right)\
,
\end{eqnarray}
where the parameter $c=14.4$ is related to a magnetic cutoff (see
Ref.~\cite{Peshier}).

In line with Ref.~\cite{DQPM3}, the parton spectral functions are no
longer $\delta$ functions in the invariant mass squared but have a
Lorentzian form,
\begin{eqnarray}
\label{20} \rho_{j}(\omega,{\bf
p}\!)=\!\frac{\Gamma_{j}}{E_j}\!
\left(\frac{1}{(\omega-E_j)^2+\Gamma^{2}_{j}}
-\frac{1}{(\omega+E_j)^2+\Gamma^{2}_{j}}\right) \end{eqnarray}
$$
=\frac{4\omega\Gamma_j}{(\omega^2-{\bf
p}^2-M_j^2)^2+4\Gamma^2_j\omega^2}, $$
\\[0.1cm]
with the notation $E_{j}^2({\bf p}^2)={\bf
p}^2+M_{j}^{2}-\Gamma_{j}^{2}$, where the index $j$ stands for quarks,
antiquarks and gluons ($j = q,\bar q,g$). The spectral function
(\ref{20}) is antisymmetric in $\omega$ and normalized as
\begin{equation}
\int\limits_{-\infty}^{\infty}\frac{d\omega}{2\pi}\
\omega\rho_{j}(\omega,{\bf p})=
\int\limits_{0}^{\infty}\frac{d\omega}{2\pi}\ 2\omega
\rho_{j}(\omega,{\bf p})=1\ .
\end{equation}
The parameters $\Gamma_{j}$ and $M_{j}$ from the DQPM have been
defined above. Note, however, that the decomposition of the total
width $\Gamma_j$ into the collisional width (due to elastic and
inelastic collisions) and the decay width is not addressed in the
DQPM. The effective cross sections for each of the various partonic
channels as a function of the energy density $\varepsilon$, which
fixes the partial widths of the dynamical quasiparticles as well as
the various interaction rates, have been  determined in
Ref.~\cite{box}.

\subsection{Hadronization in PHSD}
\label{hadronization} The hadronization, i.e., the transition from
partonic to hadronic degrees of freedom and vice versa, is described
in PHSD by covariant transition rates for the fusion of
quark-antiquark pairs to mesonic resonances or three quarks
(antiquarks) to baryonic states \cite{PHSD2}, e.g., for $q+\bar{q}$
fusion to a meson $m$ of four-momentum $p= (\omega, {\bf p})$ at
space-time point $x=(t,{\bf x})$:
\begin{eqnarray}
&&\hspace{-0.45cm} \frac{d N_m(x,p)}{d^4x d^4p}\!=\! {\rm Tr}_q
{\rm Tr}_{\bar q}\delta^4(p\!-\!p_q\!-\!p_{\bar q})\delta^4\!\!\left(\frac{x_q+x_{\bar q}}{2}-x\right) \nonumber\\
&&\hspace{0.9cm}\times\omega_q\rho_{q}(p_q)\omega_{\bar
q}\rho_{{\bar q}}(p_{\bar
q})|v_{q\bar{q}}|^2W_m\!\left(x_q-x_{\bar q},\frac{p_q-p_{\bar q}}{2}\right)\nonumber \\
&&\hspace{0.9cm}\times N_q(x_q, p_q)N_{\bar q}(x_{\bar q},p_{\bar
q})\delta({\rm flavor},{\rm color})\ . \label{trans}
\end{eqnarray}
In Eq.~\eqref{trans} we have introduced the shorthand notation,
\begin{equation}
{\rm Tr}_j = \sum_j \int d^4x_j \int \frac{d^4p_j}{(2\pi)^4} \ ,
\end{equation}
where $\sum_j$ denotes a summation over discrete quantum numbers
(spin, flavor, color); $N_j(x,p)$ is the phase-space density of
parton $j$ at space-time position $x$ and four-momentum $p$. In
Eq.~\eqref{trans} $\delta({\rm flavor},\, {\rm color})$ stands
symbolically for the conservation of flavor quantum numbers as well
as color neutrality of the formed hadron $m$, which can be viewed as
a color-dipole or ``prehadron.''  Furthermore, $v_{q{\bar
q}}(\rho_p)$ is the effective quark-antiquark interaction  from the
DQPM (displayed in Fig.~10 of Ref. \cite{DQPM2}) as a function of
the local parton ($q + \bar{q} +g$) density $\rho_p$ (or energy
density). Furthermore, $W_m(x,p)$ is the dimensionless phase-space
distribution of the formed prehadron; i.e.,
\begin{equation}
\label{Dover} W_m(\xi,p_\xi) = \exp\left( \frac{\xi^2}{2 b^2}
\right)\exp\left[ 2 b^2 \left(p_\xi^2- \frac{(M_q-M_{\bar
q})^2}{4}\right) \right]
\end{equation}
\\[0.1cm]
with $\xi = x_1-x_2 = x_q - x_{\bar q}$ and $p_\xi = (p_1-p_2)/2 =
(p_q - p_{\bar q})/2$ (which had been introduced in
Ref.~\cite{Dover}). The width parameter $b$ has been fixed by
$\sqrt{\langle r^2 \rangle} = b$ = 0.66 fm (in the rest frame),
which corresponds to an average rms radius of mesons. We note that
the expression (\ref{Dover}) corresponds to the limit of independent
harmonic oscillator states and that the final hadron-formation rates
are approximately independent of the parameter $b$ within reasonable
variations. By construction the quantity (\ref{Dover}) is Lorentz
invariant; in the limit of instantaneous ``hadron formation,'' i.e.,
$\xi^0=0$, it provides a Gaussian dropping in the relative distance
squared $({\bf r}_1 - {\bf r}_2)^2$. The four-momentum dependence
reads explicitly (except for a factor $1/2$)
\begin{equation} (E_1 - E_2)^2 - ({\bf p}_1 - {\bf p}_2)^2 -
(M_1-M_2)^2 \leq 0,
\end{equation}
and leads to a negative argument of the second exponential in
Eq.~\eqref{Dover} favoring the fusion of partons with low relative
momenta $p_q - p_{\bar q}= p_1-p_2$.

Note that, due to the off-shell nature of both partons and hadrons,
the hadronization process obeys all conservation laws (i.e., the
four-momentum conservation and the flavor current conservation) in
each event, the detailed balance relations, and the increase in the
total entropy $S$ for rapidly expanding systems. The physics behind
Eq.~\eqref{trans} is that the inverse reaction, i.e., the
dissolution of hadronic states to quark-antiquark pairs (in the case
of mesons), at low energy density is inhibited by the large masses
of the partonic quasiparticles according to the DQPM. Vice versa the
resonant $q-{\bar q}$ pairs have a large phase-space to decay to
several $0^-$ octet mesons. We recall that the transition matrix
element becomes huge below the critical energy density \cite{PHSD2}.
For further details on the PHSD off-shell transport approach and
hadronization we refer the reader to Refs.
\cite{PHSD1,PHSD2,Cassing,Bratkovskaya:2011wp,box}.

\section{Calculation of shear and bulk viscosity coefficients}
\begin{figure}
\includegraphics[width=0.5\textwidth]{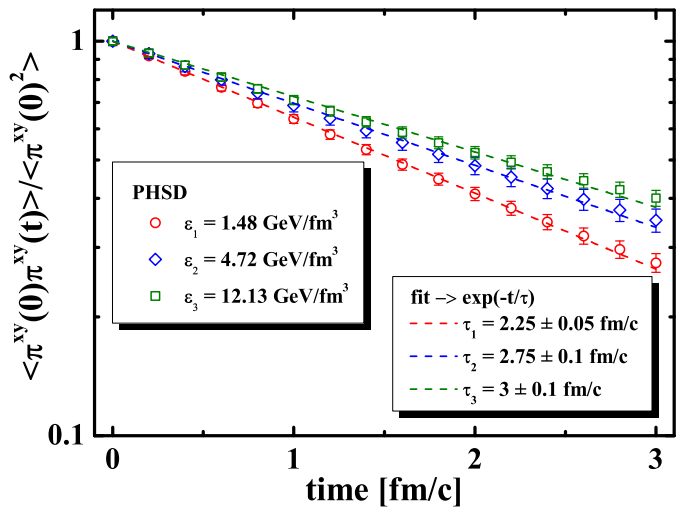}
\caption{(Color online) The correlation functions
$\bigl\langle\pi^{xy}(0)\pi^{xy}(t)\bigr\rangle$, which are
normalized by $\bigl\langle\pi^{xy}(0)^2\bigr\rangle$, as a function
of time from the PHSD simulations in the box (open symbols) for
systems at different energy densities. The corresponding exponential
fits are given by dashed lines; the extracted relaxation times
$\tau$ are given too.}\label{correlator}
\end{figure}

\begin{figure}
\includegraphics[width=0.5\textwidth]{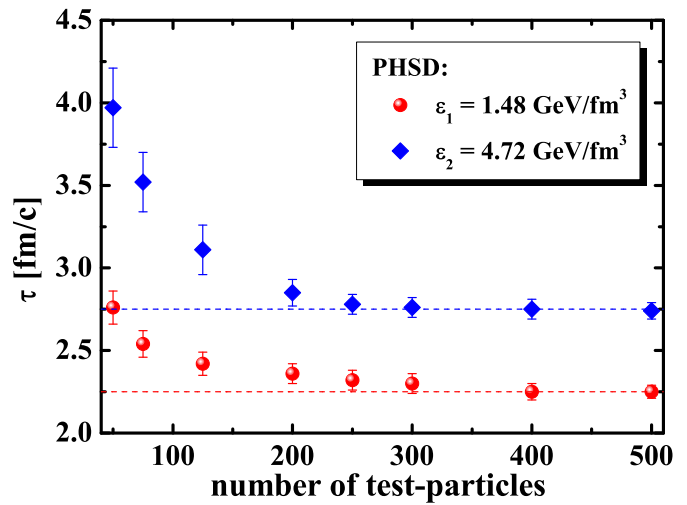}
\caption{(Color online) The relaxation time $\tau$ as a function of
the number of testparticles obtained by the PHSD simulations in the
box (symbols) for systems at different energy densities. The dashed
lines provide the convergent values for the relaxation time
$\tau$.}\label{tau}
\end{figure}

In this section we concentrate on the extraction of the shear and
bulk viscosities for ``infinite'' parton-hadron matter employing the
Green-Kubo formalism and the RTA. We simulate the ``infinite''
matter within a cubic box with periodic boundary conditions at
various values for the energy density within PHSD. The size of the
box is fixed to $9^3$ fm$^3$. The initialization is done by
populating the box with light ($u,d$) and strange ($s$) quarks,
antiquarks, and gluons. If the energy density in the system is below
the critical energy density ($\varepsilon_c \approx $ 0.5
GeV/fm$^3$), the evolution proceeds through the dynamical phase
transition (as described in Sec.~\ref{hadronization}) and ends up in
an ensemble of interacting hadrons. The system is initialized
slightly out of equilibrium and, at all energy-densities, approaches
kinetic and chemical equilibrium during it's evolution within PHSD
as was shown in our previous investigations in Ref. \cite{box}.
After equilibration, the properties of the system at given
temperature $T$ can be studied. For more details we refer the reader
to Ref.~\cite{box}, where the particle abundances, spectra,
fluctuations, and spectral functions have been studied. In the
present work we extend our investigations to the calculation of
transport coefficients.

\subsection{The Kubo formalism}
The Kubo formalism relates linear transport coefficients such as
heat conductivity and shear and bulk viscosities to nonequilibrium
correlations of the corresponding dissipative fluxes and treats
dissipative fluxes as perturbations to local thermal equilibrium
\cite{Green,Kubo}. The Green-Kubo formula for the shear viscosity
$\eta$ is as follows \cite{GKformula}:
\begin{equation} \label{kubo1}
\eta=\frac{1}{T}\int d^3r\int\limits_{0}^{\infty}
dt\bigl\langle\pi^{xy}({\bf 0},0)\pi^{xy}({\bf r},t)\bigr\rangle,
\end{equation}
where $T$ is the temperature of the system and  $\langle...\rangle$
denotes the ensemble average in thermal equilibrium. In
Eq.~(\ref{kubo1}), $\pi^{xy}$ is the shear component (nondiagonal
spacial part) of the energy momentum tensor $\pi^{\mu\nu}$:
\begin{equation}
\pi^{xy}({\bf r},t)\equiv T^{xy}({\bf
r},t)=\int\frac{d^3p}{(2\pi)^3}\frac{p^{x}p^{y}}{E}f({\bf r} ,{\bf
p};t),
\end{equation}
where the scalar mean-field $U_s$ (from PHSD) enters in the energy
$E=\sqrt{{\bf p}^2+U_s^2}$.

\begin{figure}
\includegraphics[width=0.5\textwidth]{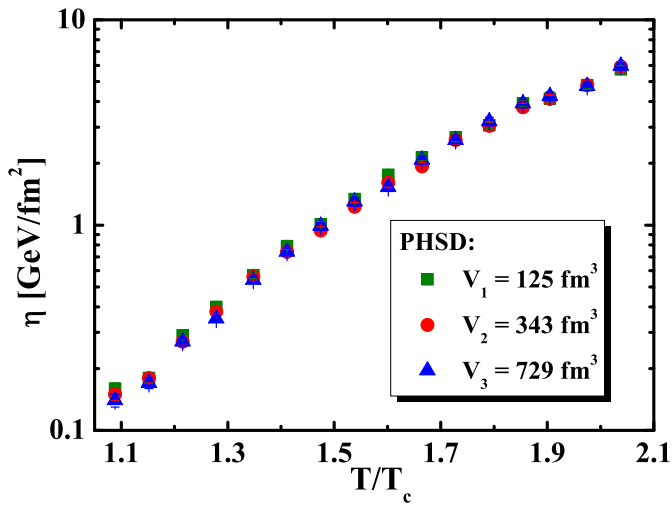}
\caption{(Color online) The shear viscosity $\eta$ as a function of
temperature   from the PHSD simulations in the box for various
volumes of the box: $V=125$ GeV/fm$^3$ (green squares), $V=343$
GeV/fm$^3$ (red circles), and $V=729$ GeV/fm$^3$ (blue
triangles).}\label{volume}
\end{figure}

In our numerical simulation---within the testparticles
representation---the volume averaged shear component of the energy
momentum tensor can be written as
\begin{equation}
\pi^{xy}(t)=\frac{1}{V}\sum\limits_{i=1}^{N}\frac{p_i^xp_i^y}{E_i},
\end{equation}
where $V$ is the volume of the system and the sum is over all
particles in the box at time $t$. Note that the scalar mean-field
contribution $U_s$ only enters via the energy $E$. Taking into
account that point particles are uniformly distributed in our box
[implying $\pi^{xy}({\bf r},t)=\pi^{xy}(t)$], we can simplify the
Kubo formula for the shear viscosity to
\begin{equation}
\eta=\frac{V}{T}\int\limits_{0}^{\infty}dt\bigl\langle\pi^{xy}(0)\pi^{xy}(t)\bigr\rangle.
\end{equation}
The correlation functions
$\bigl\langle\pi^{xy}(0)\pi^{xy}(t)\bigr\rangle$ are empirically
found to decay almost exponentially in time,
\begin{equation}
\bigl\langle\pi^{xy}(0)\pi^{xy}(t)\bigr\rangle=\bigl\langle\pi^{xy}(0)\pi^{xy}(0)\bigr\rangle\
e^{-t/\tau},
\end{equation}
as shown in Fig.~\ref{correlator}, where $\tau$ is the respective
relaxation time. Finally, we end up with the Green-Kubo formula for
the shear viscosity:
\begin{equation}
\eta=\frac{V}{T}\bigl\langle\pi^{xy}(0)^2\bigr\rangle\tau,
\end{equation}
which we use to extract the shear viscosity from the PHSD
simulations in the box at given energy density. Note that the
temperature $T$ is uniquely related to the energy density
$\varepsilon(T)$ in PHSD (in thermodynamic equilibrium).

We check the numeric stability of the method by plotting the
respective relaxation times $\tau$, extracted from the PHSD
simulations in the box, as a function of the number of testparticles
in Fig.~\ref{tau}. The results for the relaxation time $\tau$
converge for $N_{test}\geqslant400$ independent of the energy
density. In this study, we use  a high amount of microcanonical
simulations in the ensemble average ($N_{test}=500$), which leads to
reliable (within statistical error bars) results.

We also note that our numerical results for $\eta$ do not depend on
the volume $V$ of the box within reasonable variations by factors of
6 as shown in Fig.~\ref{volume}.

\subsection{The relaxation time approximation}

The starting hypothesis of the RTA is that the collision integral
can be approximated by
\begin{equation}
C[f]=-\frac{f-f^{eq}}{\tau},
\end{equation}
where $\tau$ is the relaxation time. In this approach it has been
shown that the shear and bulk viscosities (without mean-field or
potential effects) can be written as \cite{Hosoya,Gavin,Kapusta}
\begin{equation}
\eta=\frac{1}{15T}\sum\limits_{a}\int\frac{d^3p}{(2\pi)^3}\frac{|{\bf
p}|^4}{E_a^2}\tau_a(E_a)f^{eq}_a(E_a/T),
\end{equation}
\begin{equation}
\zeta=\frac{1}{9T}\sum\limits_{a}\int\frac{d^3p}{(2\pi)^3}
\frac{\tau_a(E_a)}{E_a^2}\bigl[(1-3v_s^2)E_a^2-m_a^2\bigr]^2 \ f^{eq}_a(E_a/T),
\end{equation}
where the sum is over particles of different type $a$ (in our case,
$a=q,\bar q,g$). In the PHSD transport approach the relaxation time
is given by
\begin{equation}
\tau_a(T)=\Gamma^{-1}_a(T),
\end{equation}
where $\Gamma_a(T)$ is the width of particles of type $a=q,\bar q,g$
as defined by Eqs.~(\ref{widthg}) and (\ref{widthq}).  In our
numerical simulation---within the testparticle representation
---the volume averaged shear and bulk viscosities are given by the
following expressions:
\begin{equation}
\eta=\frac{1}{15TV}\sum\limits_{i=1}^{N}\frac{|{\bf
p}_i|^4}{E_i^2}\Gamma^{-1}_i,
\end{equation}
\begin{equation}
\zeta=\frac{1}{9TV}\sum\limits_{i=1}^{N}\frac{\Gamma^{-1}_i}{E_i^2}\bigl[(1-3v_s^2)E_i^2-m_i^2\bigr]^2,
\end{equation}
where the speed of sound $v_s=v_s(T)$ is taken from lQCD \cite{lQCD}
or the DQPM, alternatively. Note that $v_s(T)$ from both approaches
is practically identical since it is governed by the DQPM, which
reproduces the lQCD results.

\begin{figure}
\includegraphics[width=0.5\textwidth]{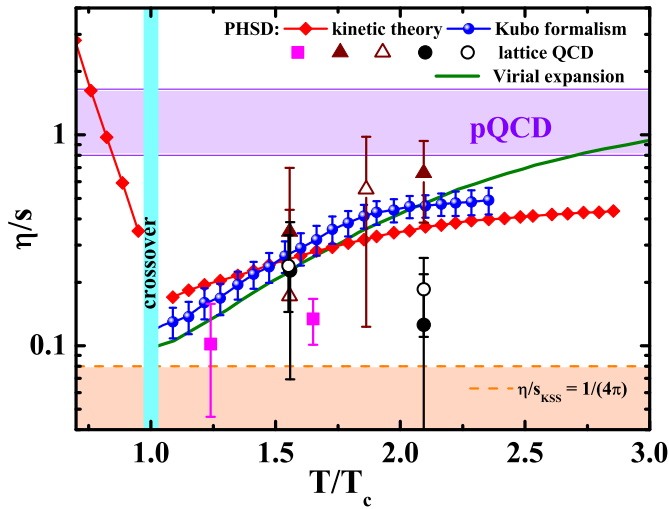}
\caption{(Color online) The shear viscosity to entropy density ratio
$\eta/s$ as a function of temperature of the system obtained by the
PHSD simulations using different methods: the RTA (red
line$+$diamonds) and the Kubo formalism (blue line$+$dots). The
other symbols denote lQCD data for pure $SU_c(3)$ gauge theory from
Ref.~\cite{lQCDtransport1} (magenta squares), from
Ref.~\cite{lQCDtransport3} (open and solid triangles), and from
Ref.~\cite{lQCDtransport4} (black open and solid circles). The
orange dashed line demonstrates the Kovtun-Son-Starinets bound
\cite{KSS} $(\eta/s)_{KSS}=1/(4\pi).$ For comparison, the results
from the virial expansion approach (green line) \cite{Mattiello} are
shown as a function of temperature too.}\label{shear}
\end{figure}

\subsection{Results for the shear viscosity}
In Fig.~\ref{shear} we present the shear viscosity to entropy
density ratio $\eta/s$ as a function of temperature $T$ of the
system extracted from the PHSD simulations in the box, where the
viscosity was extracted employing the RTA(red line$+$diamonds) and
the Kubo formalism (blue line$+$dots). We find that these approaches
give roughly the same $\eta/s$ as a function of temperature within
error bars. For comparison, the results from the virial expansion
approach~\cite{Mattiello} are given by the green line as well as
lQCD data for pure $SU_c(3)$ gauge theory.  The results for $T <
T_c$ stem from PHSD in the relaxation time framework and rapidly
rise with decreasing temperature. This is mainly because of a strong
decrease of the entropy density, $s\rightarrow0$ at $T\rightarrow0$
as $e^{-m_{\pi}/T}$.

The behavior of the specific shear viscosity with temperature in
PHSD is in agreement with the results of the scaling hadron masses
and couplings and ``heavy quark bag'' (SHMC-HQB)
approach~\cite{Khvorostukhin1,Khvorostukhin2,Khvorostukhin:2012kw},
where the partonic phase is described in the ``heavy quark bag''
model.
However, we obtain considerably lower values for the shear
viscosity, in particular, in the partonic phase. The low viscosity
of the quark-gluon matter in PHSD is caused by the stronger
interaction between the degrees of freedom  and is supported by the
successful description of experimental data on the collective flow
in heavy-ion collisions within PHSD~\cite{PHSD2,Koncha11}.

At $T<T_c$, the PHSD results for the viscosity of the hadronic
matter at vanishing quark chemical potential $\mu_q=0$
qualitatively agree with the calculations in
Refs.~\cite{Chen:2006iga,Gorenstein:2007mw,Lang:2012tt,NoronhaHostler:2012ug}.
On the other hand, let us note that the results for the hadronic
phase here have to be extended to finite $\mu_q$ before
applications to realistic heavy-ion collisions can be performed.
This is the topic of a separate forthcoming study.

\subsection{Mean-field or potential effects}
We recall that for vanishing quark chemical potential the partonic
mean fields are essentially of scalar type and vector or tensor
fields are suppressed, since the average quark current is zero.
Furthermore, partonic mean fields affect the bulk viscosity but not
the shear viscosity (except for a contribution in the energy $E$ in
the denominator). According to  Ref.~\cite{Kapusta}, the expression
for the bulk viscosity with potential effects reads
\begin{eqnarray}
\nonumber\zeta
&=&\frac{1}{T}\sum\limits_{a}\int\frac{d^3p}{(2\pi)^3}\frac{\tau_a(E_a)}{E_a^2}f^{eq}_a(E_a/T)\\
&\times &\Bigl[\Bigl(\frac{1}{3}-v_s^2\Bigr)|{\bf
p}|^2-v_s^2\Bigl(m_a^2-T^2\frac{dm_a^2}{dT^2}\Bigr)\Bigr]^2.
\end{eqnarray}
In the numerical simulation the volume averaged bulk viscosity
(including the mean-field effects from PHSD) is evaluated as
\begin{equation} \label{bulk3}
\zeta=\frac{1}{TV}\sum\limits_{i=1}^{N}\frac{\Gamma^{-1}_i}{E_i^2}\Bigl[\Bigl(\frac{1}{3}-v_s^2\Bigr)|{\bf
p}|^2-v_s^2\Bigl(m_i^2-T^2\frac{dm_i^2}{dT^2}\Bigr)\Bigr]^2.
\end{equation}
By using the DQPM expressions for the masses of quarks and gluons
(for $\mu_q=0$),
$$m_q^2(T/T_c)=\frac{1}{3}g^2(T/T_c)T^2,\,\,\,\,m_g^2(T/T_c)=\frac{3}{4}g^2(T/T_c)T^2,$$
we can calculate the derivatives with respect to $T^2$. Thus all
quantities in Eq.~(\ref{bulk3}) are uniquely determined within PHSD.
We recall that the DQPM description of thermodynamic properties of
lQCD results \cite{lQCD} and its implementation in PHSD give
practically the same results \cite{box}. The derivation of partonic
mean fields as well as their values can be found in
Ref.~\cite{DQPM1}.

\begin{figure}
\includegraphics[width=0.5\textwidth]{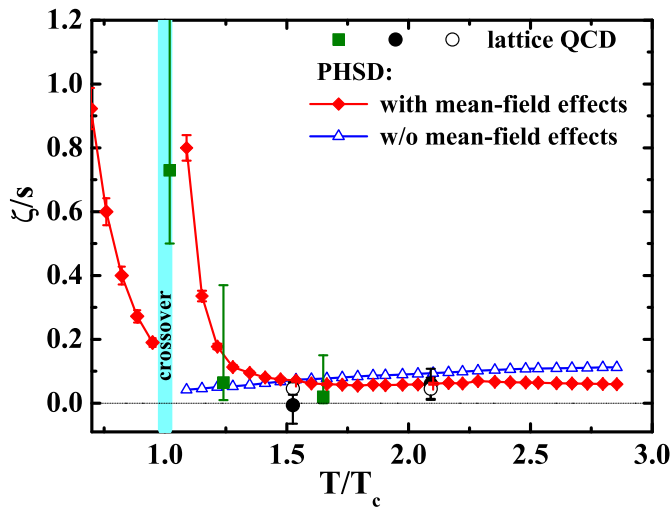}
\caption{(Color online) The bulk viscosity to entropy density ratio
$\zeta/s$ as a function of temperature $T$  extracted from the PHSD
simulations in the box using the RTA with mean-field effects (red
line$+$diamonds) and without potential effects (blue line$+$open
triangles). The available lQCD data from Ref.~\cite{lQCDtransport2}
are given by green squares and from Ref.~\cite{lQCDtransport4} by
black open and solid circles, respectively.}\label{bulk}
\end{figure}

\subsection{Results for the bulk viscosity}
In Fig.~\ref{bulk} we show the bulk viscosity to entropy density
ratio $\zeta/s$  as a function of temperature $T$ of the system
obtained by the PHSD simulations in the box employing the RTA with
mean-field (or potential) effects (red line$+$diamonds) and without
potential effects (blue line$+$open triangles) for the partons. For
comparison, we show in the same figure the available lQCD
data~\cite{lQCDtransport2,lQCDtransport4}. Without mean-field
effects we find an almost constant ratio $\zeta (T)/\eta (T)$ (see
below), which is not in line with the findings from the lattice.
Thus the dynamical mean fields (as incorporated in PHSD) play a
decisive role in the temperature dependence of the bulk viscosity
$\zeta(T)$ of the strongly interacting quark-gluon plasma. The
increase of the bulk viscosity per unit entropy at $T\approx T_c$ is
generated by the collective interaction of partons via mean fields
rather than by their scatterings. At high temperature the mean-field
effects are less pronounced and the values for the bulk viscosity of
partonic matter from PHSD are approaching those obtained in the
scope of the SHMC-HQB
model~\cite{Khvorostukhin1,Khvorostukhin2,Khvorostukhin:2012kw}.

\begin{figure}
\includegraphics[width=0.5\textwidth]{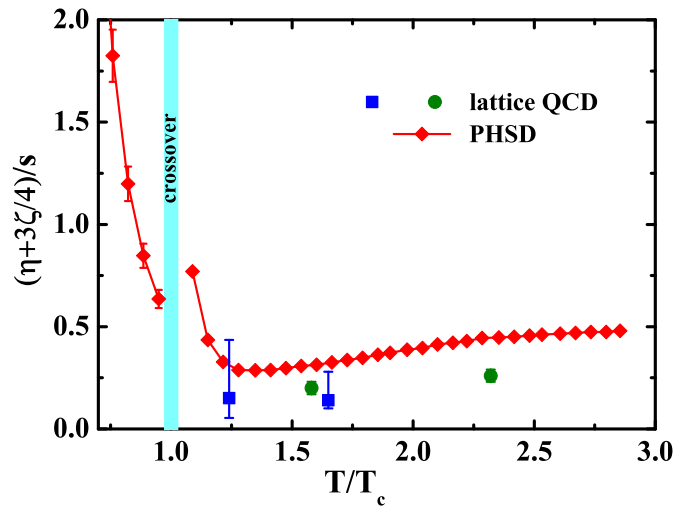}
\caption{(Color online) The specific sound channel
$(\eta+3\zeta/4)/s$ as a function of temperature $T$  of the system
obtained by the PHSD simulations in the box using the RTA with
mean-field effects (red line$+$diamonds). It is compared with lQCD
data from Ref.~\cite{lQCDtransport5} (green circles) and from
combining the results of Refs.~\cite{lQCDtransport2} and
\cite{lQCDtransport4} (blue squares).}\label{sound_channel}
\end{figure}

On the hadronic side, we observe that $\zeta/s$ falls with
temperature, which is in agreement with the results of the SHMC–-HQB
model~\cite{Khvorostukhin1,Khvorostukhin2,Khvorostukhin:2012kw} and
of the chiral model for an interacting pion
gas~\cite{Dobado:2011qu,Lu:2011df}. However, we do not see a
divergent behavior of the bulk viscosity to entropy density ratio
for $T \to 0$ as predicted in Ref.~\cite{Lu:2011df}.

Further related quantities are of interest, in particular, the
specific sound $(\eta+3\zeta/4)/s$. A sound wave propagation in the
$z$ direction with wavelength $\lambda=2\pi/k$ is damped according
to
\begin{equation}
T_{03}(t,k)\propto\exp{\Biggl[-\frac{\bigl(\frac{4}{3}\eta+\zeta\bigr)k^2t}{2(\varepsilon+p)}\Biggr]},
\end{equation}
where $T_{03}$ is the momentum density in the $z$ direction,
$\varepsilon$ is the energy density, and $p$ is the pressure. Thus
both the shear $\eta$ and bulk $\zeta$ viscosities contribute to the
damping of sound waves in the medium and provide a further
constraint on the viscosities.
In Fig.~\ref{sound_channel} we present the specific sound channel
$(\eta+3\zeta/4)/s$ as a function of temperature $T$ of the system
obtained by the PHSD simulations in the box using the RTA with
mean-field effects (red line$+$diamonds). It is compared with lQCD
results for pure $SU_c(3)$ gauge theory from
Ref.~\cite{lQCDtransport5} (green circles) and from combining the
results of Refs.~\cite{lQCDtransport1} and \cite{lQCDtransport2}
(blue squares). Note that the PHSD calculations correspond to
unquenched three-flavor QCD and thus are not expected to match the
results for the pure gauge theory exactly.

\begin{figure}
\includegraphics[width=0.5\textwidth]{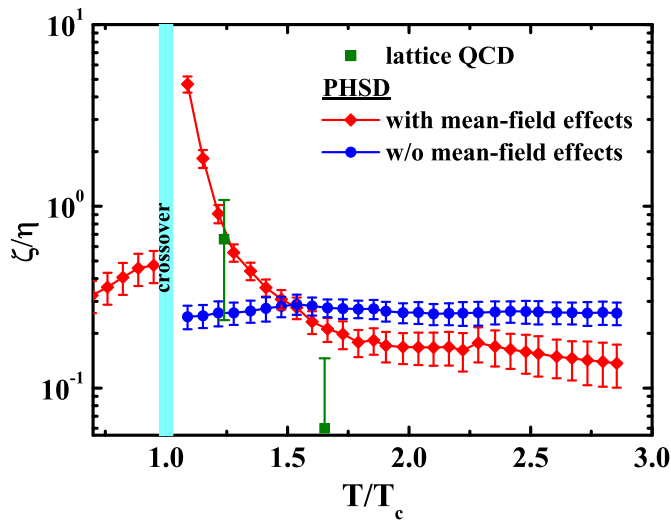}
\caption{(Color online) The bulk to shear viscosity ratio
$\zeta/\eta$ as a function of temperature of the system obtained by
the PHSD simulations in the box employing the RTA with mean-field
effects (red line$+$diamonds) and without potential effects (blue
line$+$circles). It is compared with lQCD data from
Refs.~\cite{lQCDtransport1,lQCDtransport2} (green squares). Note the
logarithmic scale in $\zeta/\eta$. }\label{ratio}
\end{figure}

Finally, in Fig.~\ref{ratio}, we show the bulk to shear viscosity
ratio $\zeta/\eta$ as a function of temperature of the system
extracted from the PHSD simulations in the box using the RTA with
mean-field (or potential) effects (red line$+$diamonds) and without
potential effects (blue line$+$circles). Whereas an almost
temperature-independent result is obtained in the partonic phase
when discarding mean-field effects, a strong increase close to $T_c$
is found in the PHSD when including the mean fields for the partons.
The results for the shear to bulk viscosity ratio in the deconfined
phase are in agreement with the lattice
data~\cite{lQCDtransport1,lQCDtransport2} and with
Ref.~\cite{Bluhm:2011xu}. Since the PHSD gives a minimum in the
shear viscosity $\eta$ and a strong maximum in the bulk viscosity
$\zeta$ close to $T_c$ (note the logarithmic scale), the ratio
$\zeta/\eta$ has a sizable maximum in the area of the (crossover)
phase transition.

\section{Summary and conclusions}
We have employed the off-shell PHSD approach in a finite box with
periodic boundary conditions for the study of the shear and bulk
viscosities as a function of temperature (or energy density) for
dynamical infinite partonic and hadronic systems in equilibrium. The
PHSD transport model is based on a lQCD equation of state
\cite{lQCD} and well describes the entropy density $s(T)$, the
energy density $\varepsilon(T)$, and the pressure $p(T)$ in
thermodynamic equilibrium in comparison to the lQCD results
\cite{PHSD1,PHSD2,box}. We have employed the Kubo formalism as well
as the RTA to calculate the shear viscosity $\eta(T)$.   We find
that both methods provide very similar results for the ratio
$\eta/s$ with a minimum close to the critical temperature $T_c$
while approaching the perturbative QCD limit at higher temperatures.
For $T < T_c$, i.e., in the hadronic phase, the ratio $\eta/s$ rises
fast with decreasing temperature due to a lower interaction rate of
the hadronic system and a significantly smaller number of degrees of
freedom (or entropy density). Our results are, furthermore, also in
almost quantitative agreement with the ratio $\eta (T)/s (T)$ from
the virial expansion approach in Ref. \cite{Mattiello} as well as
with lQCD data for the pure gauge sector.

We have, furthermore, evaluated the bulk viscosity $\zeta (T)$ in
the RTA and focused on the effects of mean fields (or potentials) in
the partonic phase. Here we find a significant rise of the ratio
$\zeta(T)/s (T)$ in the vicinity of the critical temperature $T_c$
due to the scalar mean fields from PHSD. The result for this ratio
is in line with that from lQCD calculations. Additionally, the
specific sound $(\eta + 3 \zeta/4)/s (T)$ has been calculated and
presents a nontrivial temperature dependence; the absolute value for
this combination of the shear and bulk viscosities is in an
approximate agreement with the lattice gauge theory. Furthermore,
the ratio $\zeta(T)/\eta(T)$ within the PHSD calculations shows a
strong maximum close to $T_c$, which has to be attributed to
mean-field (or potential) effects that in PHSD are encoded in the
infrared enhancement of the resummed coupling $g(T)$.

Because the PHSD calculations have proven to describe
single-particle as well as collective observables from relativistic
nucleus-nucleus collisions from lower SPS to top RHIC energies, the
extracted transport coefficients $\eta(T)$ and $\zeta(T)$ are
compatible with experimental observations in a wide
energy/temperature range. Furthermore, the qualitative and partly
quantitative agreement with lQCD results is striking.

\section*{Acknowledgements}
The authors appreciate fruitful discussions with A.~Merdeev,
V.~Skokov, and G.~Torrieri. V.O. acknowledges financial support
through the HIC for FAIR framework of the LOEWE Program and H-QM
Graduate School. O.L. acknowledges financial support through the
Margarete-Bieber Program of the Justus-Liebig-University of Giessen.
The work of M.I.G. was supported by the Humboldt Foundation and the
Program of Fundamental Research of the Department of Physics and
Astronomy of NAS, Ukraine.


\begin{thebibliography}{99}
%
\bibitem{StrCoupled1}
M.~Gyulassy and L.~D.~McLerran, Nucl. Phys. A {\bf 750}, 30 (2005).
%
\bibitem{StrCoupled2}
E.~V.~Shuryak, Nucl. Phys. A {\bf 750}, 64 (2005).
%
\bibitem{StrCoupled3}
U.~W.~Heinz, nucl-th/0407114.
%
\bibitem{Peshier}
A.~Peshier and W.~Cassing, Phys. Rev. Lett. {\bf 94}, 172301 (2005).
%
\bibitem{STAR}
J.~Adams {\it et al.} (STAR Collaboration), Nucl. Phys. A {\bf 757},
102 (2005).
%
\bibitem{PHENIX}
K.~Adcox {\it et al.} (PHENIX Collaboration), Nucl. Phys. A {\bf
757}, 184 (2005).
%
\bibitem{BRAHMS}
I.~Arsene {\it et al.} (BRAHMS Collaboration), Nucl. Phys. A {\bf
757}, 1 (2005).
%
\bibitem{PHOBOS}
B.~B.~Back {\it et al.} (PHOBOS Collaboration), Nucl. Phys. A {\bf
757}, 28 (2005).
%
\bibitem{ALICE}
K.~Aamodt {\it et al.} (ALICE Collaboration), Phys. Rev. Lett. {\bf
105}, 252302 (2010).
%
\bibitem{IdealHydro1}
P.~Huovinen, P.~F.~Kolb, U.~W.~Heinz, P.~V.~Ruuskanen, and
S.~A.~Voloshin, Phys. Lett. B {\bf 503}, 58 (2001).
%
\bibitem{IdealHydro2}
P.~F.~Kolb, P.~Huovinen, U.~Heinz, and H.~Heiselberg, Phys. Lett. B
{\bf 500}, 232 (2001).
%
\bibitem{IdealHydro3}
D.~Teaney, J.~Lauret, and E.~V.~Shuryak, Phys. Rev. Lett. {\bf 86},
4783 (2001).
%
\bibitem{IdealHydro4}
T.~Hirano and K.~Tsuda, Phys. Rev. C {\bf 66}, 054905 (2002).
%
\bibitem{IdealHydro5}
P.~F.~Kolb and R.~Rapp, Phys. Rev. C {\bf 67}, 044903 (2003).
%
\bibitem{IdealHydro6}
P.~Huovinen, in {\it Quark-Gluon Plasma 3}, edited by R.~C.~Hwa and
X.-N.~Wang (World Scientific, Singapore, 2004); P.~F.~Kolb and
U.~W.~Heinz, edited by R.~C.~Hwa and X.-N.~Wang (World Scientific,
Singapore, 2004).
%
\bibitem{NonzeroViscosity}
P.~Danielewicz and M.~Gyulassy, Phys. Rev. D {\bf 31}, 53 (1985).
%
\bibitem{KSS}
G.~Policastro, D.~T.~Son, A.~O.~Starinets, Phys. Rev. Lett. {\bf
87}, 081601 (2001); P.~K.~Kovtun, D.~T.~Son, A.~O.~Starinets, Phys.
Rev. Lett. {\bf 94}, 111601 (2005).
%
\bibitem{Buchel}
A.~Buchel, Phys. Lett. B {\bf 663}, 286 (2008).
%
\bibitem{ViscousHydro1}
P.~Romatschke, U.~Romatschke, Phys. Rev. Lett. {\bf 99}, 172301
(2007).
%
\bibitem{ViscousHydro2}
H.~Song and U.~W.~Heinz, Phys. Rev. C {\bf 77}, 064901 (2008).
%
\bibitem{ViscousHydro3}
M.~Luzum and P.~Romatschke, Phys. Rev. C {\bf 78}, 034915 (2008).
%
\bibitem{ViscousHydro4}
B.~Schenke, S.~Jeon, and C.~Gale, Phys. Rev. C {\bf 82}, 014903
(2010).
%
\bibitem{Minshear}
L.~P.~Csernai, J.~I.~Kapusta, and L.~D.~McLerran, Phys. Rev. Lett.
{\bf 97}, 152303 (2006).
%
%
\bibitem{Mattiello}
S.~Mattiello and W.~Cassing, Eur. Phys. J. C \textbf{70}, 243
(2010).
%
\bibitem{Toneev}
A.~S.~Khvorostukhin, V.~D.~Toneev, and D.~N.~Voskresensky, Phys.
Rev. C \textbf{83}, 035204 (2011).
%
\bibitem{MaxBulk1}
D.~Kharzeev and K.~Tuchin, JHEP \textbf{09}, 093 (2008).
%
\bibitem{MaxBulk2}
F.~Karsch, D.~Kharzeev, and K.~Tuchin, Phys. Lett. B {\bf 663}, 217
(2008).
%
\bibitem{MaxBulk3}
P.~Romatschke and D.~T.~Son, Phys. Rev. D {\bf 80}, 065021 (2009).
%
\bibitem{MaxBulk4}
G.~D.~Moore and O.~Saremi, JHEP {\bf 09}, 015 (2008).
%
\bibitem{MaxBulk5}
C.~Sasaki and K.~Redlich, Phys. Rev. C {\bf 79}, 055207 (2009);
Nucl. Phys. A {\bf 832}, 62 (2010).
%
\bibitem{Pratt}
K.~Paech and S.~Pratt, Phys. Rev. C {\bf 74}, 014901 (2006).
%
\bibitem{Giorgio}
G.~Torrieri and I.~Mishustin, Phys. Rev. C {\bf 78}, 021901 (2008).
%
\bibitem{pQCD1}
P.~B.~Arnold, G.~D.~Moore, and L.~G.~Yaffe, JHEP {\bf 11}, 001
(2010); {\bf 05}, 051 (2003).
%
\bibitem{pQCD2}
P.~B.~Arnold, C.~Dogan, and G.~D.~Moore, Phys. Rev. D {\bf 74},
085021 (2006).
%
\bibitem{pQCD3}
M.~Prakash, M.~Prakash, R.~Venugopalan, and G.~Welke, Phys. Rep.
{\bf 227}, 321 (1993).
%
\bibitem{pQCD4}
J.~W.~Chen and J.~Wang, Phys. Rev. C {\bf 79}, 044913 (2009).
%
\bibitem{lQCDtransport1}
H.~B.~Meyer, Phys. Rev. D {\bf 76}, 101701 (2007).
%
\bibitem{lQCDtransport2}
H.~B.~Meyer, Phys. Rev. Lett. {\bf 100}, 162001 (2008).
%
\bibitem{lQCDtransport3}
A.~Nakamura and S.~Sakai, Phys. Rev. Lett. {\bf 94}, 072305 (2005).
%
\bibitem{lQCDtransport4}
S.~Sakai and A.~Nakamura, Pos {\bf LAT2007}, 221 (2007).
%
\bibitem{RTA}
F.~Reif, {\it Fundamentals of Statistical and Thermal Physics}
(McGraw-Hill, New York, 1965).
%
\bibitem{CE}
S.~R.~de~Groot, W.~A.~van~Leeuwen, and C.~Weert, {\it Relativistic
Kinetic Theory, Principles and Applications}, North-Holland Company,
Amsterdam, 1980.
%
\bibitem{Green}
M.~S.~Green, J. of Chem. Phys. {\bf 22}, 398 (1954).
%
\bibitem{Kubo}
R.~Kubo, J. Phys. Soc. Japan {\bf 12}, 570 (1957); Rep. Prog. Phys.
{\bf 29}, 255 (1966).
%
\bibitem{Bluhm}
M.~Bluhm, B.~K\"ampfer, and K.~Redlich, Phys. Rev. C {\bf 84},
025201 (2011).
%
\bibitem{GrecoQGP}
S.~Plumari, A.~Puglisi, F.~Scardina, and V.~Greco, nucl-th/12080481.
%
\bibitem{Plumari}
S.~Plumari, W.~M.~Alberico, V.~Greco, and C.~Ratti, Phys. Rev. D
{\bf 84}, 094004 (2011).
%
\bibitem{Thoma}
M.~H.~Thoma, Phys. Lett. B {\bf 269}, 144 (1991).
%
\bibitem{Khvorostukhin1}
A.~S.~Khvorostukhin, V.~D.~Toneev, and D.~N.~Voskresensky, Nucl.
Phys. A {\bf 845}, 106 (2010).
%
\bibitem{Khvorostukhin2}
A.~S.~Khvorostukhin, V.~D.~Toneev and D.~N.~Voskresensky, Phys. Rev.
C {\bf 84}, 035202 (2011).
%
\bibitem{Muronga}
A.~Muronga, Phys. Rev. C {\bf 69}, 044901 (2004).
%
\bibitem{Bass}
N.~S.~Demir, S.~A.~Bass, Eur. Phys. J. C {\bf 62}, 63 (2009).
%
\bibitem{Pal}
S.~Pal, Phys. Lett. B {\bf 684}, 211 (2010).
%
\bibitem{Wesp}
C.~Wesp, A.~El, F.~Reining, Z.~Xu, I.~Bouras, and C.~Greiner, Phys.
Rev. C {\bf 84}, 054911 (2011).
%
\bibitem{BassQGP}
J.~Fuini~III, N.~S.~Demir, D.~K.~Srivastava, and S.~A.~Bass, J.
Phys. G {\bf 38}, 015004 (2011).
%
\bibitem{PHSD1}
W.~Cassing and E.~L.~Bratkovskaya, Nucl. Phys. A \textbf{831}, 215
(2009).
%
\bibitem{PHSD2}
W.~Cassing and E.~L.~Bratkovskaya, Phys. Rev. C \textbf{78}, 034919
(2008).
%
\bibitem{Olena}
O.~Linnyk, E.~L.~Bratkovskaya, V.~Ozvenchuk, W.~Cassing and
C.~M.~Ko, Phys. Rev. C {\bf 84}, 054917 (2011).
%
\bibitem{Olena2}
O.~Linnyk, W.~Cassing, J.~Manninen, E.~L.~Bratkovskaya and C.~M.~Ko,
Phys. Rev. C {\bf 85}, 024910 (2012).
%
\bibitem{Koncha11}
V.~P.~Konchakovski {\it et al.}, Phys. Rev. C \textbf{85}, 044922
(2012); J. Phys. Conf. Ser. \textbf{389}, 012015 (2012).
%
\bibitem{Kadanoff1}
L.~P.~Kadanoff and G.~Baym, {\it Quantum Statistical Mechanics},
(Benjamin, New York, 1962).
%
\bibitem{Kadanoff2}
S.~Juchem, W.~Cassing, and C.~Greiner, Phys. Rev. D \textbf{69},
025006 (2004); Nucl. Phys. A \textbf{743}, 92 (2004).
%
\bibitem{CBRep98}
W. Cassing and E. L. Bratkovskaya, Phys. Rept. {\bf 308}, 65 (1999).
%
\bibitem{Brat97}
E.~L.~Bratkovskaya and W.~Cassing, Nucl. Phys. A {\bf 619}, 413
(1997).
%
\bibitem{Cass90} W. Cassing,  V. Metag, U. Mosel, and K. Niita,
Phys. Rep. \textbf{188}, 363 (1990).
%
\bibitem{DQPM1}
W.~Cassing, Nucl. Phys. A \textbf{795}, 70 (2007).
%
\bibitem{DQPM2}
W.~Cassing, Nucl. Phys. A \textbf{791}, 365 (2007).
%
\bibitem{DQPM3}
A.~Peshier, Phys. Rev. D \textbf{70}, 034016 (2004); J. Phys. G
\textbf{31}, S371 (2005).
%
\bibitem{lQCD}
Y. Aoki {\it et al.}, Phys. Lett. B \textbf{643}, 46 (2006); S. Borsanyi {\it et al.},
JHEP \textbf{1009} 073 (2010).
%
\bibitem{Bratkovskaya:2011wp}
E.~L.~Bratkovskaya, W.~Cassing, V.~P.~Konchakovski, and O.~Linnyk,
Nucl. Phys. A \textbf{856}, 162 (2011).
%
\bibitem{Pisarski}
R.~D.~Pisarski, Phys. Rev. Lett. {\bf 63}, 1129 (1989).
%
\bibitem{box}
V.~Ozvenchuk, O.~Linnyk, M.~I.~Gorenstein, E.~L.~Bratkovskaya, and
W.~Cassing, Phys. Rev. C {\bf 87}, 024901 (2013).
%
\bibitem{Dover}
C.~B.~Dover, U.~Heinz, E.~Schnedermann, and J.~Zimanyi, Phys. Rev. C
{\bf 44}, 1636 (1991).
%
\bibitem{Cassing}
W.~Cassing, Eur. Phys. J. ST {\bf 168}, 3 (2009).
%
\bibitem{GKformula}
R.~Zubarev, O.~Morozov, {\it Statistical Mechanics of Nonequilibrium
Processes Volume 2: Relaxation and Hydrodynamic Processes},
(Akademie Verlaf GmbH, 1996).
%
\bibitem{Hosoya}
A.~Hosoya and K.~Kajantie, Nucl. Phys. B {\bf 250}, 666 (1985).
%
\bibitem{Gavin}
S.~Gavin, Nucl. Phys. A {\bf 435}, 826 (1985).
%
\bibitem{Kapusta}
P.~Chakraborty and J.~I.~Kapusta, Phys. Rev. C {\bf 83}, 014906
(2011).
%
\bibitem{Khvorostukhin:2012kw}
A.~S.~Khvorostukhin, V.~D.~Toneev and D.~N.~Voskresensky,
arXiv:1204.5855 [nucl-th].
%
\bibitem{Chen:2006iga}
J.-W.~Chen and E.~Nakano, Phys. Lett. B {\bf 647}, 371 (2007).
%
\bibitem{Gorenstein:2007mw}
M.~I.~Gorenstein, M.~Hauer and O.~N.~Moroz, Phys. Rev. C {\bf 77},
024911 (2008).
%
\bibitem{Lang:2012tt}
R.~Lang, N.~Kaiser and W.~Weise, Eur. Phys. J. A {\bf 48}, 109
(2012).
%
\bibitem{NoronhaHostler:2012ug}
J.~Noronha-Hostler, J.~Noronha and C.~Greiner, Phys. Rev. C {\bf
86}, 024913 (2012).
%
\bibitem{Dobado:2011qu}
A.~Dobado, F.~J.~Llanes-Estrada and J.~M.~Torres-Rincon, Phys. Lett.
B {\bf 702}, 43 (2011).
%
\bibitem{Lu:2011df}
E.~Lu and G.~D.~Moore, Phys. Rev. C {\bf 83}, 044901 (2011).
%
\bibitem{lQCDtransport5}
H.~B.~Meyer, Nucl. Phys. A {\bf 830}, 641c (2009).
%
\bibitem{Bluhm:2011xu}
M.~Bluhm, B.~K\"ampfer and K.~Redlich, Phys. Lett. B {\bf 709}, 77
(2012).

\end{thebibliography}
\end{document}